\documentclass[journal,transmag]{IEEEtran}

\ifCLASSINFOpdf
\else
\fi
\usepackage{graphicx}

\begin{document}

\title{Perpendicular standing spin wave and magnetic anisotropic study on amorphous FeTaC films}

\author{\IEEEauthorblockN{B.~ Samantaray\IEEEauthorrefmark{1},
Akhilesh Kr.~ Singh\IEEEauthorrefmark{2},
A.~ Perumal\IEEEauthorrefmark{2}, and
P.~ Mandal\IEEEauthorrefmark{1}}
\IEEEauthorblockA{\IEEEauthorrefmark{1}Saha Institute of Nuclear Physics, 1/AF Bidhannagar, Calcutta 700064, India}
\IEEEauthorblockA{\IEEEauthorrefmark{2}Department of Physics, Indian Institute of Technology Guwahati, Guwahati 781039, India}
\thanks{Manuscript received November 7, 2015; revised September 00, 0000.
Corresponding author: B. Samantaray (email: iitg.biswanath@gmail.com).}}


\IEEEtitleabstractindextext{%
\begin{quote}
\begin{abstract}
Magnetic anisotropy, spin wave (SW) excitation and exchange stiffness constant of amorphous FeTaC ($d$ = 20-200 nm) films were studied as a function of thickness using micro-strip ferromagnetic resonance (MS-FMR) technique. The MS-FMR spectra for in-plane applied magnetic field show the presence of uniform precessional mode ($n$ = 0) along with first perpendicular standing spin wave (PSSW) mode ($n$ = 1) especially for $d$ = 50, 100 and 200 nm films. The angular ($\varphi_{H}$) dependence of resonance field ($H_{r}$) and magnetic field dependence of resonance frequencies ($f_{r}$) in planar configuration for the uniform and PSSW modes were modeled successfully by using dispersion relation which arises from a combination of exchange and dipolar interactions. The relevant parameters such as saturation magnetization ($4\pi M_{S}$), uniaxial anisotropic constant ($K_{u}$), $g$-factor, and exchange stiffness constants ($A_{ex}$) are estimated for different FeTaC film thickness. $A_{ex}$ is found to increase from 1.52(4)$\times$10$^{-7}$ to 5.0(5)$\times$10$^{-6}$ erg/cm as the thickness of film increases from 50 to 200 nm, possibly due to surface pinning effect or significant inhomogeneity especially at higher thickness films.
\end{abstract}
\end{quote}

}

\maketitle

\IEEEdisplaynontitleabstractindextext

\IEEEpeerreviewmaketitle

\section{INTRODUCTION}
\IEEEPARstart{S}{oft} ferromagnetic (FM) alloys have been promising candidates towards potential technological applications in various magnetoelectronic devices like magnetoresistive random access memories (MRAM), magnetic tunnel junctions (MTJs) and soft underlayer in perpendicular magnetic recording media.  The amorphous nature of those alloys reduces the number of pinning centers which may lead to the spin transfer torque (STT)-driven domain wall motion along with high tunneling magnetoresistance ratio (TMR) \cite{Fukami2009}. It has been reported that the addition of 20 $\%$ metalloid (in this case 'Ta' and 'C') in Fe based FM metal matrix destroys crystallinity and shows enhanced soft magnetic properties \cite{Miura1996,Hindmarch2008,Perumal2009,Akhilesh2012,Akhilesh2014}. Furthermore, the soft magnetism in Fe based nanocrystalline alloys arises from two-phase microstructure in which fine nanocrystals are embedded in amorphous matrix, resulting a strong intergranular FM exchange coupling \cite{Herzer1990}. Therefore, FeTaC soft FM thin films are considered to be as one of the potential candidates for applications. FeTaX (X = N or C) based nanocrystalline thin films exhibit the features of core materials in magnetic reading heads \cite{Okumura1992,Goto1994}.  In order to use this kind of alloy for development of spin devices, it is required to understand the spin wave excitations.

Ferromagnetic resonance (FMR) technique is one of the powerful tools to explore spin wave resonance and magnetization dynamics in ferromagnets with finite size. In FMR, the external microwave field couples with uniform and non uniform spin wave modes. The modes with wave vector ($\vec{q}$) component perpendicular to the film surface is termed as perpendicular standing spin wave (PSSW) modes and the quantization of these modes is due to the pinning of the moments at the surface \cite{Gui2007}. In the current work, the magnetic anisotropy, spin wave excitation and exchange stiffness are explored along with thickness dependence ($d$ = 20-200 nm) study in FeTaC soft FM thin films by using mainly micro-strip ferromagnetic resonance (MS-FMR) technique.
\section{EXPERIMENTAL DETAILS}
Single layer Fe$_{80}$Ta$_{8}$C$_{12}$ films with different thickness  $d$ = 20, 50, 100, and 200 nm were deposited by dc magnetron sputtering technique and the details were reported elsewhere \cite{Akhilesh2012}. The static and dynamic magnetic properties were explored by using a vector network analyzer (VNA) based custom made MS-FMR spectrometer and vibrating sample magnetometer (LakeShore model no. 7410). The details of MS-FMR technique and measurement procedure are discussed elsewhere \cite{Samantaray2015}. The magnetic field sweep FMR spectra have been carried out for different precessional frequencies and azimuthal angles. The frequency and angular dependence of resonance fields were extracted from each FMR spectrum and the numerical modeling was performed by using mathematica program.
\section{THEORETICAL BACKGROUND}
In this section, frequency ($f$) and azimuthal angle ($\varphi_{H}$) dependence of resonance fields ($H_{r}$) will be derived for uniform precessional mode and PSSW mode in order to interpret the experimental results. The free energy density of a single magnetic thin film can be written as,
\begin{equation}
\begin{array}{c}
E =  - {M_S}H\left[ {\sin {\theta _H}\sin {\theta _M}cos\left( {{\varphi _H} - {\varphi _M}} \right) + \cos {\theta _H}\cos {\theta _M}} \right]\\
 - 2\pi M_S^2{\sin ^2}{\theta _M} - {K_u}\left( \begin{array}{l}
{\sin ^2}{\theta _M}{\cos ^2}{\varphi _M}{\cos ^2}{\varphi _u}\\
 + {\sin ^2}{\theta _M}{\sin ^2}{\varphi _M}{\sin ^2}{\varphi _u}
\end{array} \right)\\
 + {K_ \bot }{\sin ^2}{\theta _M}
\end{array}
\end{equation}
In the above expression, $\varphi_{H}$  and $\varphi_{M}$ are azimuthal angles corresponding to $H$ and $M$ directions, respectively. $\theta_{H}$ and $\theta_{M}$ are polar angles. The first term in the above expression corresponds to the Zeeman energy, and the second term is dipolar demagnetization energy, where as the third and fourth terms correspond to the uniaxial planar and perpendicular magnetic anisotropy energies, respectively. $M_{S}$ is the saturation magnetization, $K_{u}$ and $K_{\bot}$ are in-plane and out-of-plane uniaxial magnetic anisotropy constants, respectively. The resonance frequency $f_{r}$ of the uniform precession mode is deduced from the energy density by using the following expression \cite{Acher2003},
\begin{equation}
f_r^2 = {\left( {\frac{\gamma }{{2\pi }}} \right)^2}\frac{1}{{M_S^2{{\sin }^2}{\theta _M}}}\left[ {\frac{{{\partial ^2}E}}{{\partial \theta _M^2}}\frac{{{\partial ^2}E}}{{\partial \varphi _M^2}} - {{\left( {\frac{{{\partial ^2}E}}{{\partial {\theta _M}\partial {\varphi _M}}}} \right)}^2}} \right]\
\end{equation}
The resonance equations for planar configuration are solved at equilibrium position of $M$ under the applied magnetic field ($H$) by using the condition, $\frac{{\partial E}}{{\partial {\varphi _M}}}$$ = $0 and the solution of $H$ from the energy minimization condition is derived as,
\begin{equation}
H = \frac{{2{K_u}}}{{{M_S}}}\frac{{\cos {\varphi _M} sin{\varphi _M}}}{{sin({\varphi _H} - {\varphi _M})}}
\end{equation}
The in-plane ($\theta_{H}$ = $\theta_{M}$ = $\pi/2$) dispersion relation of the uniform and PSSW modes can be modeled in combined way from the total magnetic energy density which arises from the exchange and dipolar interactions and is given by \cite{Grimsditch1979,Belmeguenai2015},
\begin{equation}
{f_r} = \frac{\gamma }{{2\pi }}\left( \begin{array}{c}
{\left( \begin{array}{l}
Hcos\left( {{\varphi _M} - {\varphi _H}} \right)\\
 + \frac{{2{K_u}}}{{{M_S}}}\cos 2\left( {{\varphi _M} - {\varphi _u}} \right) + \frac{{2{A_{ex}}}}{{{M_S}}}{\left( {\frac{{n\pi }}{d}} \right)^2}
\end{array} \right)^{\frac{1}{2}}}\\
{\left( \begin{array}{l}
Hcos\left( {{\varphi _M} - {\varphi _H}} \right) + 4\pi {M_S} - \frac{{2{K_u}}}{{{M_S}}}\\
 + \frac{{2{K_u}}}{{{M_S}}}{\cos ^2}\left( {{\varphi _M} - {\varphi _u}} \right) + \frac{{2{A_{ex}}}}{{{M_S}}}{\left( {\frac{{n\pi }}{d}} \right)^2}
\end{array} \right)^{\frac{1}{2}}}
\end{array} \right)
\end{equation}
where $\gamma$ is the gyromagnetic ratio, $A_{ex}$ is the exchange stiffness constant and $n$ is the quantized number for the PSSW along the thickness direction. $n$=0 represents the uniform precession mode and the higher order modes ($n$=1, 2, 3, ...) represent PSSW mode.
\section{RESULTS AND DISCUSSION}
The room temperature magnetic hysteresis loops ($M-H$) for planar configuration are shown in Fig. 1 for different thickness of FeTaC films. The 20 nm sample shows rectangular shaped loop with remanence ratio of around 90$\%$. The loop shape is changed to flat loop along with low remanence for 50 nm sample. The coercivity fields ($H_{C}$) for 20 and 50 nm samples are found to be 1.5 and 2 Oe, respectively which is a characteristic feature of soft ferromagnetism. By increasing the film thickness ($\geq$ 100 nm), the $M-H$ clearly shows the transcritical loop manifesting the presence of stress induced perpendicular anisotropy during the film deposition \cite{Akhilesh2012}. $M_{S}$ is found to increase from 6125 $\pm$ 30 Oe to 7740 $\pm$ 40 Oe as the film thickness increases from 20 to 200 nm. The room temperature soft magnetic properties degrade drastically as thickness increases due to the transition from in-plane orientation of magnetization to the strip domain patterns \cite{Akhilesh2014}.
\begin{figure}[t]
\begin{center}
\includegraphics[trim = 0mm 0mm -5mm 0mm, width=85mm]{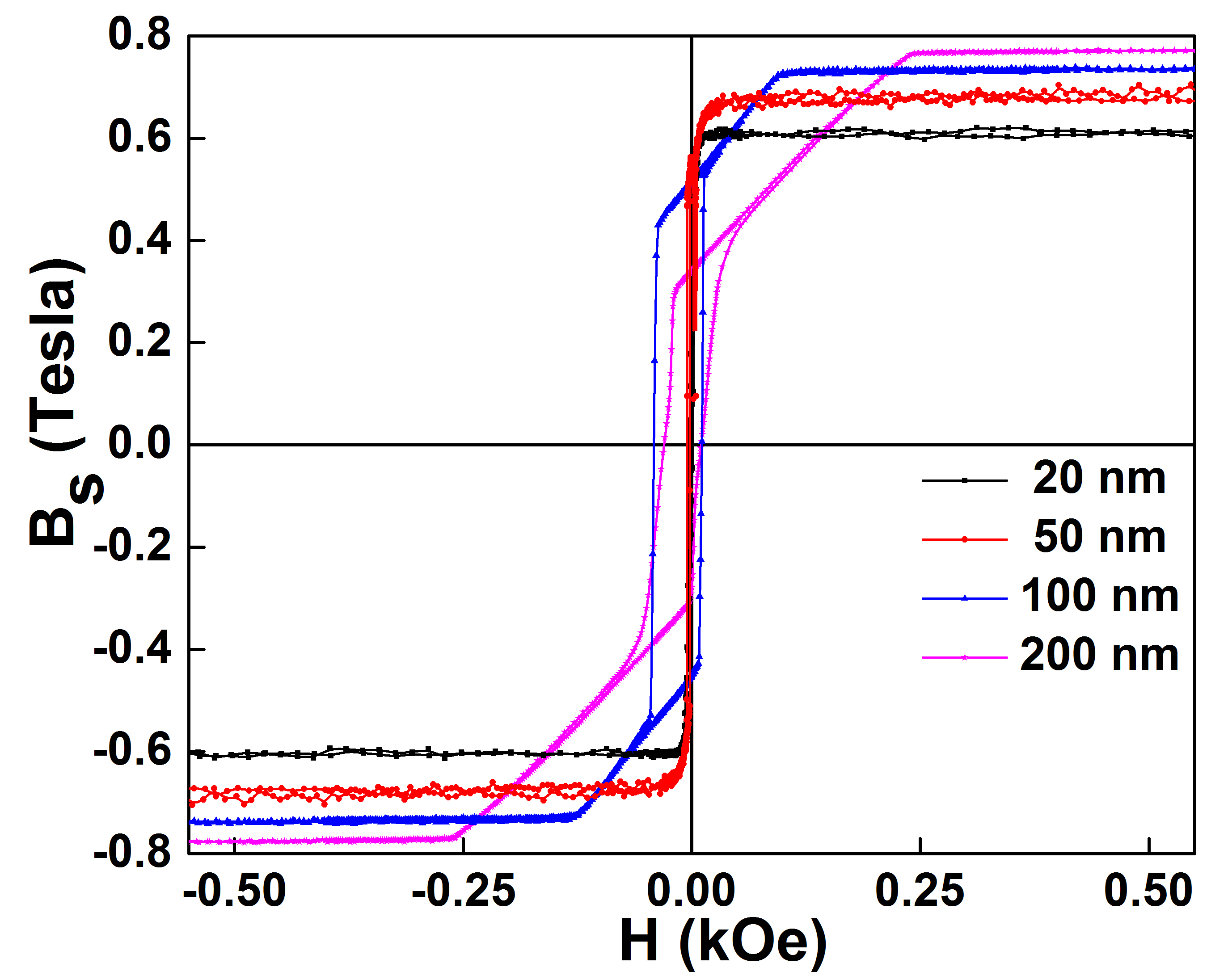}\
\caption{Room temperature $M-H$ loops for different thickness of FeTaC thin films in planar orientation.}\label{Fig.1}
\end{center}
\end{figure}
\begin{figure}[t]
\begin{center}
\includegraphics[trim = 0mm 0mm 0mm 0mm, width=75mm]{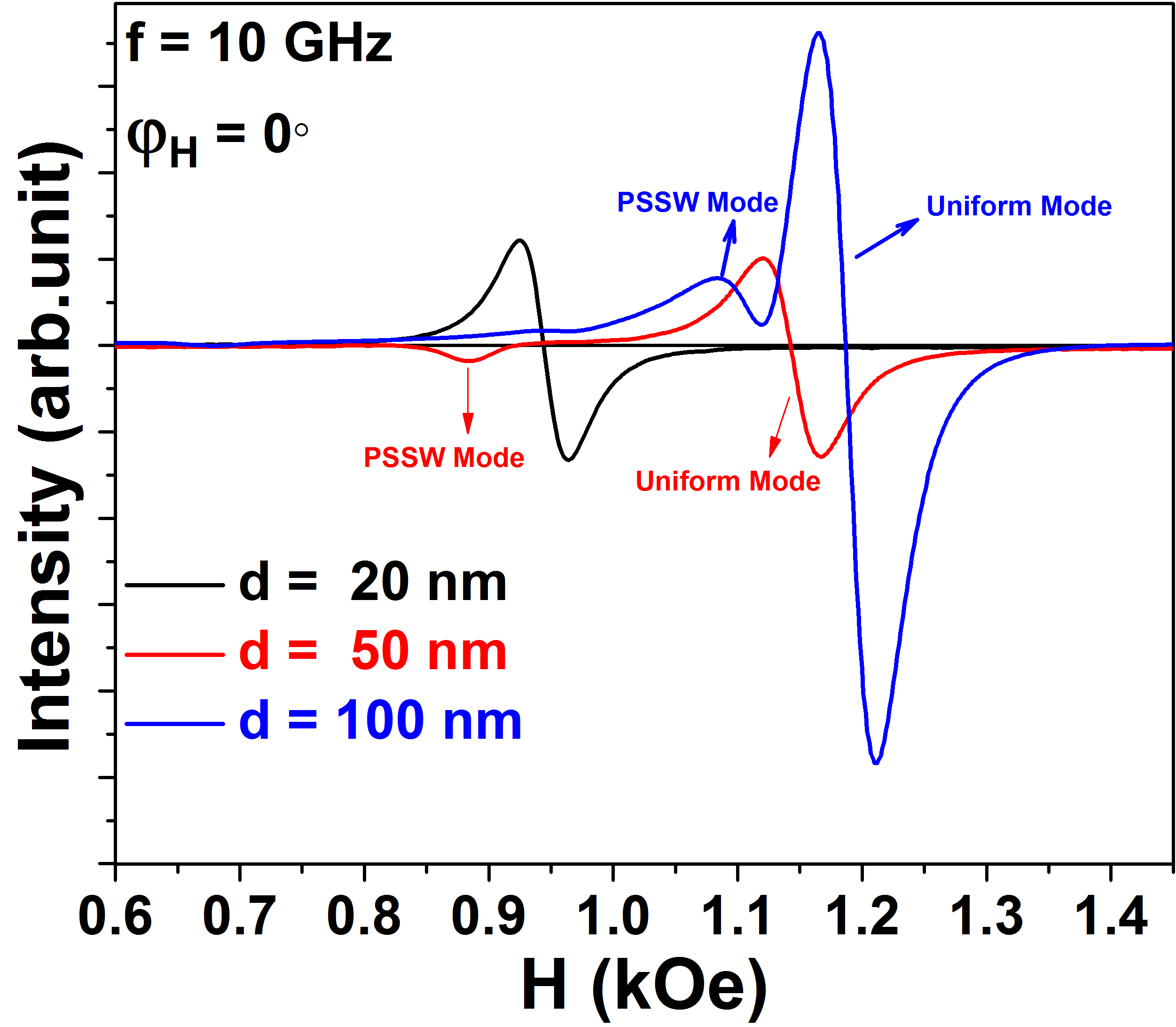}\
\caption{In-plane FMR spectra along $\varphi_{H}=0^{\circ}$ at 10 GHz for different thickness of FeTaC thin films.}\label{Fig.2}
\end{center}
\end{figure}

The typical FMR spectra for planar orientation are shown in Fig. 2 for different thickness of films at 10 GHz precessional frequency along the direction $\varphi_{H}$=0$^{\circ}$. It is observed that the 20 nm thick film shows a clear uniform precession mode without any other spin wave resonance (SWR) mode, in which the dynamic magnetization is uniform across the film thickness.
\begin{figure}[t]
\begin{center}
\includegraphics[trim = 10mm 5mm 0mm 0mm, width=95mm]{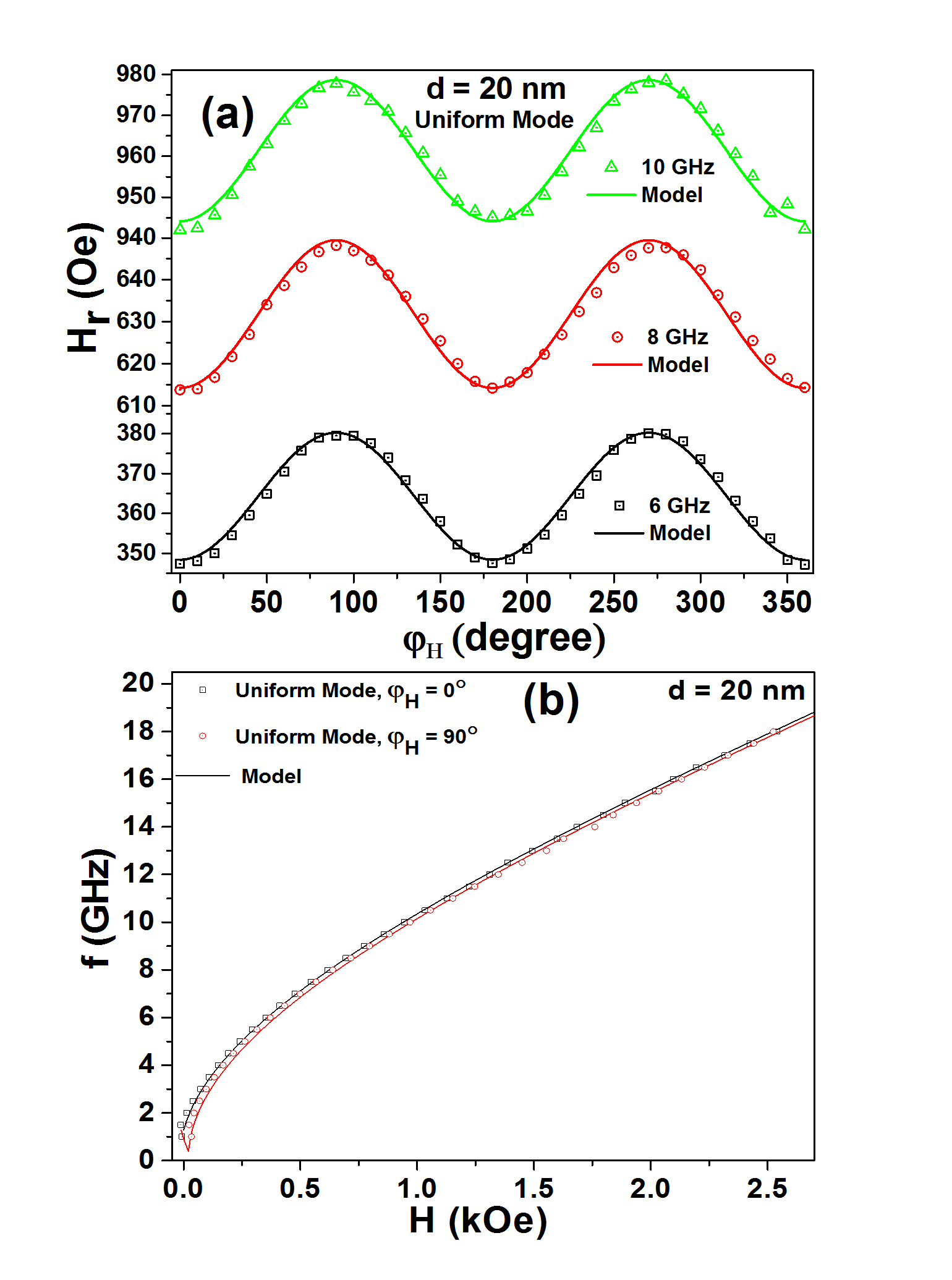}\
\caption{(a) In-plane angular dependence of the resonance fields at different precessional frequencies for d = 20 nm. (b) Resonance frequency as a function of in-plane applied magnetic field along easy and hard axis of magnetization for d = 20 nm. The solid lines are fit to the model.}\label{Fig.3}
\end{center}
\end{figure}
\begin{figure}[t]
\begin{center}
\includegraphics[trim = 10mm 10mm 0mm 0mm, width=90mm]{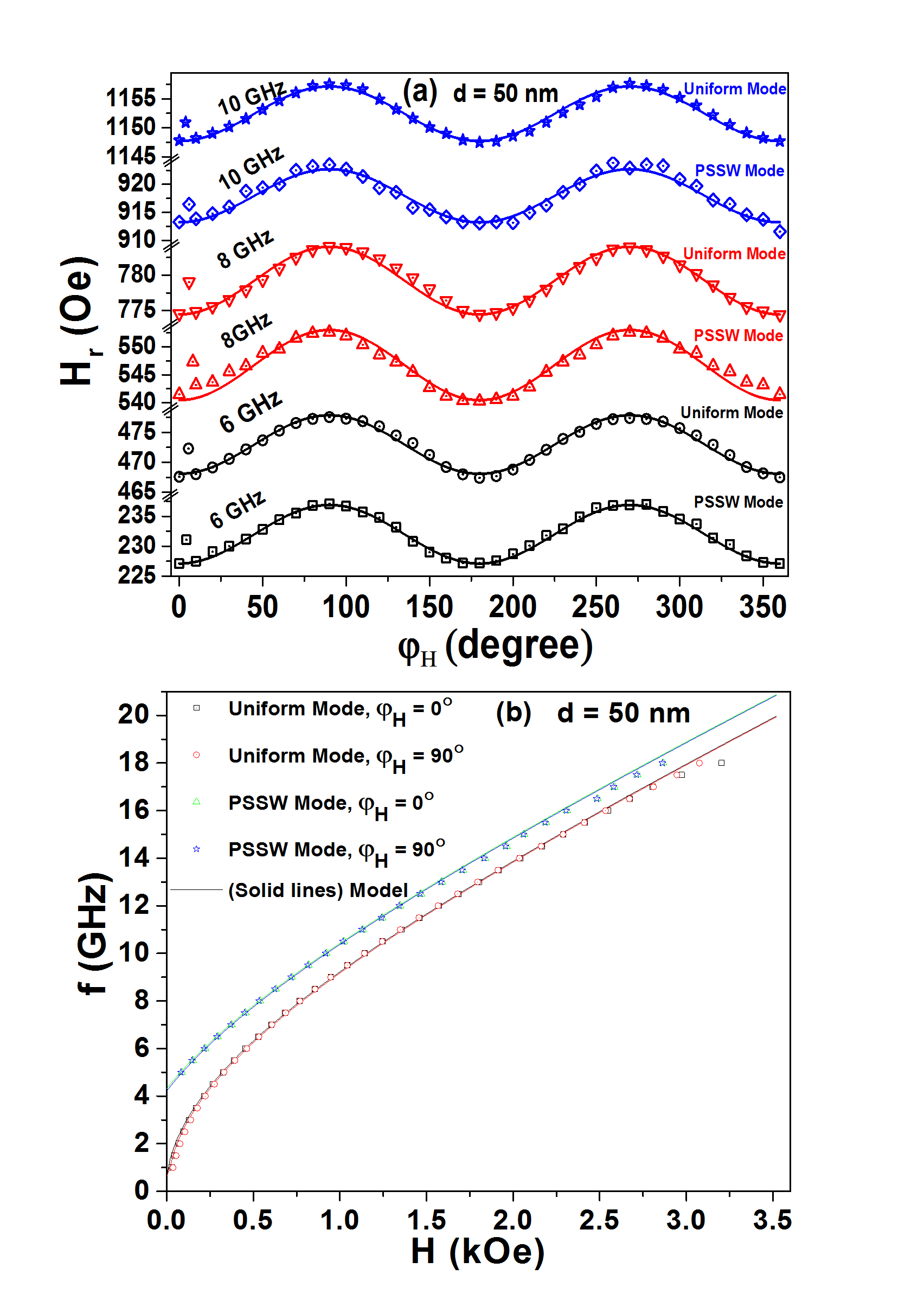}\
\caption{In-plane angular dependence of the resonance fields at different precessional frequencies for uniform and PSSW modes for d = 50 nm. Resonance frequency as a function of in-plane applied magnetic field along easy and hard axis of magnetization for uniform and PSSW modes for d = 50 nm. Experimentally and numerically calculated data points are shown as open symbols and solid lines, respectively.}\label{Fig.4}
\end{center}
\end{figure}
However, by increasing the film thickness from 50 nm onwards, the first PSSW mode is excited at lower absorption fields. The signal intensity of this mode in 50 nm thick film is 7 times smaller than the uniform mode. The absence of higher order ($n>1$) exchange-dominated PSSW modes could be due to very weak excitation and the intensity of those modes may be lower than our detection sensitivity. The separation between the uniform mode and first PSSW mode increases with decreasing the film thickness, which could be understood on the basis that the perpendicularly quantized spin wave vector is inversely proportional to the film thickness, i.e. $q=n\pi/d$. For thicker film, especially for $d$ = 200 nm, the separation between the PSSW and uniform mode is even smaller (not shown) and the broadening in uniform mode is observed along with a small kink at lower absorption field. As the thickness of the film increases, $H_{r}$ shifts towards higher absorption fields due to higher saturation magnetization. Fig. 3(a) shows the $\varphi_{H}$ dependence of $H_{r}$ for uniform mode at 6, 8, and 10 GHz precessional frequencies for 20 nm thick film, which depicts that the angular dependence of resonance fields is governed by uniaxial magnetic anisotropy.
The typical field dependence of resonance frequencies is shown in Fig. 3(b). The experimental data points are successfully modeled using Eq.(4) by incorporating the condition $n=0$. By increasing the thickness of the film from 50 nm onwards, $\varphi_{H}$ dependence of $H_{r}$ for uniform mode and the first PSSW mode at three different frequencies were modeled by incorporating the condition $n=0$ and $n=1$, respectively and are shown as solid lines in Fig. 4 (a). The respective $H$ dependence of $f_{r}$ along easy and hard axis of the magnetization is also shown in Fig. 4(b). The numerically data points yielded a good fit and the relevant parameters such as $M_{S}$, $K_{u}$, $g$-factor, and $A_{ex}$ for different film thickness are listed in Table-1. The origin of in-plane uniaxial magnetic anisotropy in FeTaC thin films ($d$ = 20 and 50 nm) could be understood on the basis that the strong exchange coupling between the FM atoms plays an important role during deposition process, which allows to form aligned FM atom pairs parallel to the film plane. The small $K_{u}$ values of 4.5(5)$\times$10$^{3}$ and 1.3(2)$\times$10$^{3}$ erg/cm$^{3}$  are estimated for 20 and 50 nm thick films, respectively and the consequent planar anisotropic fields ($H_{u}$) are 18.3 and 5 Oe. On further increase in thickness of the film from 100 to 200 nm, the in-plane uniaxial magnetic anisotropy disappears. $H$ dependence of $f_{r}$ for both the modes for 100 [Fig. 5] and 200 nm [not shown] thick films are also modeled.

\begin{table*} {
\caption{Parameters obtained from MS-FMR analysis. The abbreviations $4\pi M_{S}$, $K_{u}$, $H_{u}$, $A_{ex}$, $g$ are saturation magnetization, in-plane uniaxial magnetic anisotropy constant, in-plane uniaxial anisotropic field, gyromagnetic ratio, and exchange stiffness constant, respectively.}
\label{I}
\begin{tabular*}{1.0\textwidth}{@{\extracolsep{\fill}}c c c c c c c}
\hline 
 d & $4\pi M_{S}$ (Oe) - VSM & $4\pi M_{S}$ (Oe)- MS-FMR & $K_{u}$ (erg/cm$^{3}$) & $H_{u}$ (Oe) & $g$ & $A_{ex}$ (erg/cm) \\
\hline
20 nm & 6125$\pm$30 & 6157$\pm$5 & 4.5$\times$$10^{3}$(5) & 18.3 & 2.45(3) & $--$ \\[6pt]
50 nm & 6500$\pm$25 & 6471$\pm$5 & 1.3$\times$$10^{3}$(2) & 5 & 2.32(8) & 1.52(4)$\times$$10^{-7}$ \\[6pt]
100 nm & 7445$\pm$20 & 7414$\pm$7 & $--$ & $--$ & 2.23(1) & 2.0(2)$\times$$10^{-7}$ \\[6pt]
200 nm & 7740$\pm$40 & 7791$\pm$10 & $--$ & $--$ & 2.14(2) & 5.0(5)$\times$$10^{-6}$ \\[6pt]
\hline
\hline
\end{tabular*}}
\end{table*}

\begin{figure}[t]
\begin{center}
\includegraphics[trim = 10mm 10mm 0mm 0mm, width=70mm]{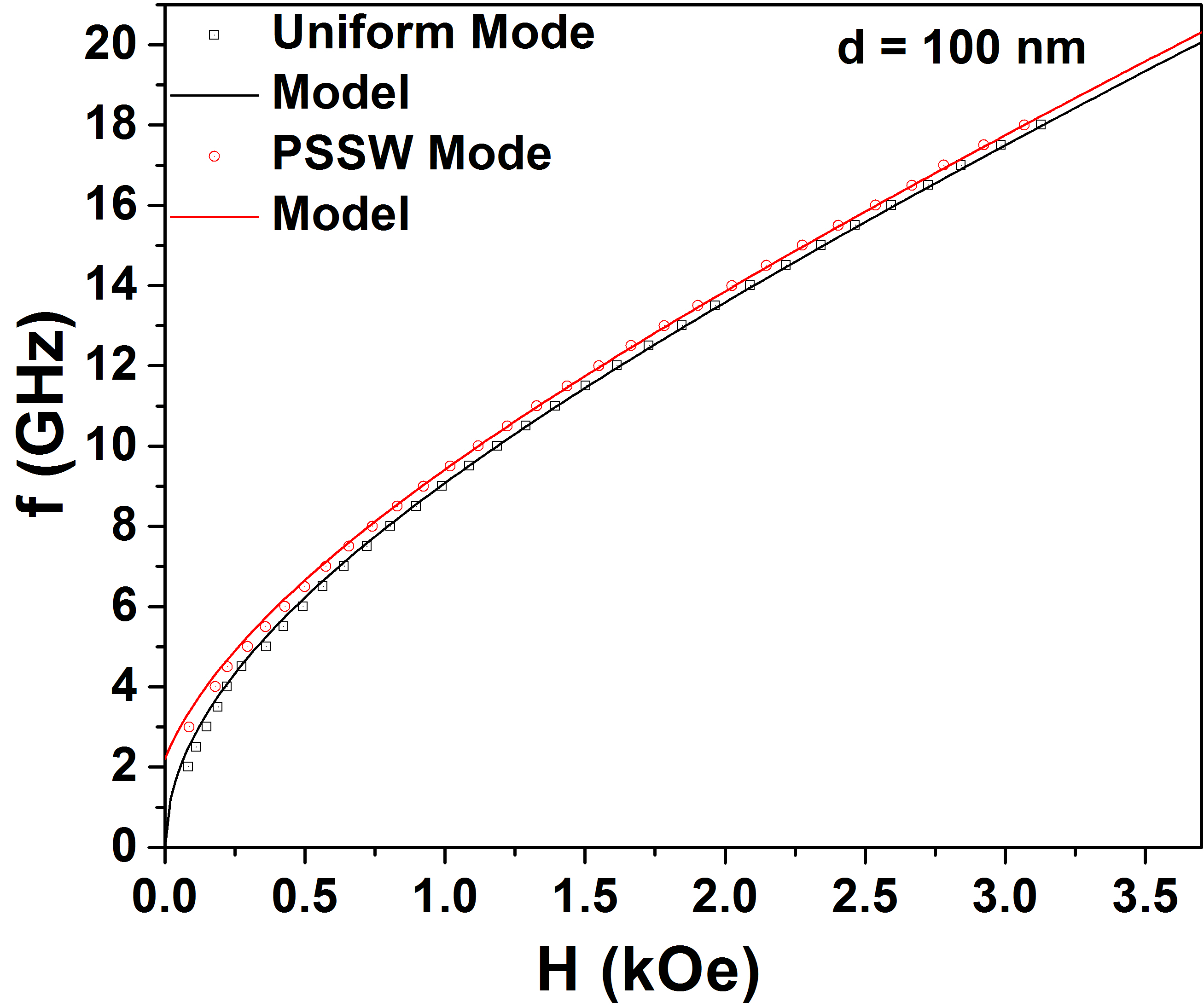}\
\caption{Resonance frequency as a function of in-plane applied magnetic field for uniform and PSSW modes for d = 100 nm. The solid lines are fit to the model.}\label{Fig.5}
\end{center}
\end{figure}

Interestingly, $M_{S}$ values obtained from MS-FMR analysis are close to that observed in $M-H$ loop measurements and are found to increase with film thickness. $g$-factor shows $1/d$ dependence behavior. $A_{ex}$ also increases with increase in thickness, which may be due to the surface pinning effect \cite{Grimsditch1979}. The values of $A_{ex}$ for 50 and 100 nm thick films are found to be one order of magnitude smaller than the known value $~1\times$10$^{-6}$ erg/cm for bulk Fe \cite{Cullity}, whereas it is comparable to FePt film of thickness 105 nm \cite{Martins2007}. On the other hand, $A_{ex}$ for 200 nm film is 5 times larger than that of Fe. The reason may be due to the significant inhomogeneity which adequately excites the spin waves or segregation of Fe clusters in such thick films.
\section{Conclusions}
MS-FMR technique has been employed to explore the magnetic anisotropy and spin wave excitations in FeTaC based soft FM films. The azimuthal angular dependence of resonance fields depicts the presence of small uniaxial magnetic anisotropy in $d$ = 20 and 50 nm thick films. The existence of first PSSW mode along with uniform mode gives rise to the evidence of exchange and dipolar interactions. The exchange stiffness constant for these amorphous films are deduced from the in-plane angular evolution as well as from the frequency spacing along the easy and hard axis of magnetization of PSSW mode.  $A_{ex}$ values are found to be increased from 1.52(4)$\times$10$^{-7}$ to 5.0(5)$\times$10$^{-6}$ erg/cm with increasing thickness of the film form 50 to 200 nm, respectively.

\end{document}